\documentclass{aa}
\usepackage{graphicx}
\usepackage{txfonts}
\usepackage{natbib}
\usepackage{aalongtable}
\bibpunct{(}{)}{;}{a}{}{,}
\newcommand{\nodata}{...}

\begin{document}

\title{Enhanced density and magnetic fields in interstellar OH masers}



\author{Vincent L.~Fish\inst{1}, Mark J.~Reid\inst{2},
Karl M.~Menten\inst{3} \& Thushara Pillai\inst{3}}

\offprints{V.~Fish, \email{vfish@nrao.edu}}

\institute{Jansky Fellow, National Radio Astronomy Observatory,
  P.~O.~Box O, 1003 Lopezville Road, Socorro, NM 87801; vfish@nrao.edu
\and
  Harvard--Smithsonian Center for Astrophysics, 60 Garden Street,
  Cambridge, MA 02138
\and
  Max-Planck-Institut f\"{u}r Radioastronomie, Auf dem H\"{u}gel 69,
  Bonn D-53121, Germany}

\date{Received date / Accepted date}

\begin{abstract}
{}
{We have observed the 6030 and 6035 MHz transitions of OH in high-mass
star-forming regions to obtain magnetic field estimates in both maser
emission and absorption.}
{Observations were taken with the Effelsberg 100 m telescope.}
{Our observations are consistent with previous results, although we do
detect a new 6030 MHz maser feature near $-70$~km~s$^{-1}$\ in the
vicinity of W3(OH).  In absorption we obtain a possible estimate of
$-1.1 \pm 0.3$~mG for the average line-of-sight component of the
magnetic field in the absorbing OH gas in K3-50 and submilligauss
upper limits for the line-of-sight field strength in DR 21 and W3.}
{These results indicate that the magnetic field strength in
the vicinity of OH masers is higher than that of the surrounding,
non-masing material, which in turn suggests that the density of masing
OH regions is higher than that of their surroundings.}
\end{abstract}

\titlerunning{Enhanced density in OH masers}



\authorrunning{Fish et al.}

\maketitle

\section{Introduction\label{intro}}

The hydroxyl radical (OH) is one of the most commonly-observed
molecules in astrophysical contexts.  OH masers are seen in massive
star-forming regions as well as other environments, such as evolved
stars and supernova remnants.  Since they are small and often bright,
their locations can be pinpointed, and multi-epoch observations
provide proper motions to submilliarcsecond astrometric accuracy.  OH
masers are also an excellent tracer of magnetic fields due to the
large Zeeman splitting coefficients in the $^2\Pi_{3/2}$ ladder, which
includes the ground-state transitions at 18 cm and the transitions in
the first rotationally excited state at 5 cm.

Our main motivation for this work is to address the question of
whether cloud regions giving rise to OH masers are denser than the
surrounding non-masing material.  If true, this suggests that maser
proper motions represent actual physical motions in the material,
rather than unrelated travelling excitation phenomena.  To this end,
we aimed to identify sources with strong OH absorption and to obtain
estimates of the magnetic field strength in the non-masing gas
therein.  The magnetic field strength is a reasonable proxy for the
density because both theoretical modelling
\citep{mouschovias76,fiedler93} and observations of molecular clouds
\citep[e.g.,][]{crutcher99} indicate that the magnetic field scales as
$n^{\kappa}$, $\kappa \lesssim 0.5$ during collapse.  This scaling
holds over many orders of magnitude in density, and the few milligauss
magnetic field strengths found from Zeeman splitting of the maser
lines correspond well to the densities ($\sim 10^6$ -- $10^7$
cm$^{-3}$) that are thought to be conducive to their excitation
\citep{cesaroni91}.

It is necessary to observe excited-state transitions of OH because the
spectra of high-mass star-forming regions with a large column density
of OH are generally dominated by maser features in the ground state.
Previous observations of main-line $^2\Pi_{3/2}, J = 7/2$ absorption
were successful in detecting the Zeeman splitting in one source
(W3(OH)) and obtaining milligauss-scale upper limits in two others
(K3-50 and G10.624$-$0.385) \citep{gusten94,uchida01,frm05}.  In this
work we extend this technique to the main-line $^2\Pi_{3/2}, J = 5/2$
transitions.  Because this is a lower excited state, the spectra of
many sources (e.g., W3(OH)) are dominated by strong masers.  However,
when uncontaminated absorption is seen, Zeeman splitting is easier to
detect, due to a larger Zeeman splitting coefficient than in the $J =
7/2$ transitions.

Several surveys of the main-line $^2\Pi_{3/2}, J = 5/2$ transitions
have been undertaken previously.  Chief among them are the survey by
\citet[][henceforth B97]{baudry97} in the northern hemisphere and a
series of surveys in the southern hemisphere by \citet{caswell95} and
\citet{caswell97,caswell01,caswell03}.  Our source list has a large
overlap with B97 and a smaller overlap with the Caswell surveys, most
of whose sources are not observable from Effelsberg, Germany.
We compare our results with these and other previous results in Sect.\
\ref{sourcenotes}.

\begin{table}
\caption{Observed Sources}
\label{source-table}
\centering
\begin{tabular}{l@{\,\,\,}l@{\,\,\,}l@{\,\,\,}r@{\,\,\,}r@{\,\,\,}l}
\hline\hline
       & RA      & Dec     & Time$^{\mathrm{a}}$ & $\sigma^{\mathrm{b}}$ & \\
Source & (J2000) & (J2000) & (min)            & (mJy) &
Detected$^\mathrm{c}$ \\
\hline
\object{W3} Cont         & 02 25 43.3 & $+$62 06 13$^\mathrm{d}$&1217  &   3 & A \\
\object{W3(OH)}          & 02 27 03.8 & $+$61 52 26 &  18 &21$^\mathrm{e}$& M       \\
\object{Orion KL}        & 05 35 14.2 & $-$05 22 47 &   6 & 216 & M?      \\
\object{S269}            & 06 14 37.1 & $+$13 49 36 &  12 &  21 & \nodata \\
\object{R Leo}           & 09 47 33.5 & $+$11 25 46 &  12 &  17 & \nodata \\
\object{IRC +10216}      & 09 47 57.3 & $+$13 16 42 &  72 &   8 & \nodata \\
\object{Sgr A}           & 17 45 40.2 & $-$29 00 30 &  12 & 107 & \nodata \\
\object{Sgr B2M}         & 17 47 20.1 & $-$28 23 22 &  84 &   8 & E M     \\
\object{G5.886$-$0.393}  & 18 00 30.6 & $-$24 04 01 &  12 &  35 & A E M\\
\object{G10.624$-$0.385} & 18 10 28.6 & $-$19 55 50 &  12 &  31 & A M  \\
\object{M17}             & 18 20 22.4 & $-$16 11 25 &   6 &\nodata$^\mathrm{f}$& M       \\
\object{G19.61$-$0.23}   & 18 27 38.1 & $-$11 56 39 &   6 &  36 & \nodata \\
\object{G28.199$-$0.048} & 18 42 58.0 & $-$04 13 58 &  30 &  12 & E M     \\
\object{G30.589$-$0.044} & 18 47 18.8 & $-$02 00 54 &   6 &  26 & \nodata \\
\object{G35.577$-$0.029} & 18 56 22.5 & $+$02 20 27 &   6 &  27 & \nodata \\
\object{W48}             & 19 01 45.5 & $+$01 13 33 &  54 &  17 & A M     \\
\object{G45.122$+$0.133} & 19 13 27.8 & $+$10 53 37 & 180 &   8 & A M     \\
\object{W51 e/d}         & 19 23 40.0 & $+$14 30 51 &  36 &  47 & A M     \\
\object{K3-50}           & 20 01 45.7 & $+$33 32 45 & 428 &   2 & A M     \\
\object{ON 1}            & 20 10 09.1 & $+$31 31 35 &  24 &  16 & M       \\
\object{ON 2 N}          & 20 21 44.0 & $+$37 26 38 &   6 &  39 & \nodata \\
\object{DR 20}           & 20 36 52.2 & $+$41 36 24 &  12 &  20 & M       \\
\object{W75 N}           & 20 38 36.4 & $+$42 37 34 &   6 &  34 & M       \\
\object{W75 S}           & 20 39 00.9 & $+$42 22 38 &  30 &  16 & M       \\
\object{DR 21}           & 20 39 01.1 & $+$42 19 43 & 640 &   3 & A       \\ 
\object{LDN 1084}        & 21 43 01.4 & $+$54 56 16 & 102 &   8 & M       \\
\object{S140}            & 22 19 18.2 & $+$63 18 46 &  30 &  13 & \nodata \\
\object{Cep A}           & 22 56 19.1 & $+$62 01 57 &   6 &  34 & \nodata \\
\object{NGC 7538}        & 23 13 45.6 & $+$61 28 18 & 180 &   8 & M       \\
\hline
\end{tabular}
\begin{list}{}{}
\item[$^{\mathrm{a}}$]Sum of on-source and off-source observing time.
\item[$^{\mathrm{b}}$]Single-polarization rms noise in a 0.243 km~s$^{-1}$
  Hanning-weighted channel.
\item[$^{\mathrm{c}}$]A = absorption, E = broad emission, M = maser emission
\item[$^{\mathrm{d}}$]Approximate position -- see Sect.\ \ref{sourcenotes}
  for details.
\item[$^{\mathrm{e}}$]Excessive ringing was still seen in 6035 MHz RCP
  after Hanning weighting.
\item[$^{\mathrm{f}}$]Autocorrelator saturated.
\end{list}
\end{table}

\section{Observations\label{observations}}

Observations were taken with the 5 cm prime focus receiver on the 100
m telescope of the Max-Planck-Institut f\"{u}r Radioastronomie
at Effelsberg between 11 and 21 March 2005.  The system temperature
was 27 K.  The gain was 1.57 K/Jy as measured using 3C286 as a flux
calibrator.  The FWHM beam size at this wavelength is approximately
130$\arcsec$.  Pointing corrections were determined only several times
a day, since pointing was stable to about 10$\arcsec$.  Focus
corrections were determined less frequently.  These corrections rarely
exceeded 1 mm, or 2\% of the observing wavelength.

The AK90 8192-channel correlator was used.  Both the $^2\Pi_{3/2}, J =
5/2, F = 3 \rightarrow 3$ (rest frequency 6035.092 MHz) and $F = 2 \rightarrow 2$
(rest frequency 6030.747 MHz) transitions were observed simultaneously
in both right (RCP) and left circular polarizations (LCP).  Source DR
21 was also briefly observed in the $F = 2 \rightarrow 3$ (rest frequency
6016.746 MHz) transition.  These transitions were Doppler-corrected
for the motion of the Earth around the Sun as well as the LSR velocity
of the source and then centered in four IFs of 5 MHz bandwidth.  Each
IF was subdivided into 2048 channels separated by 2.44 kHz,
corresponding to a channel width of 0.121~km~s$^{-1}$.  Position switching
with equal time (generally 3 min) off- and on-source was employed.

Data were reduced using \texttt{CLASS}.  Polynomial fitting was used
to subtract baselines, which were generally stable with time.
Interference was minimal; however, occasionally spurious patterns
bearing a strong resemblence to OH absorption features appeared in IF
4 (6030 MHz LCP).  These were generally easily identifiable, and the
corresponding data were kept when the velocity region of interest was
uncontaminated.  In the rare instances in which these spurious
features coincided with real absorption and/or emission features, the
scan was discarded.

Gaussian profiles were fit to the data in spectral ranes with
absorption and emission.  In most cases the data were Hanning-weighted
before fitting profiles to absorption components.  Since the resulting
channel width is comparable to a typical maser line width, Gaussian
profiles were fit to the masers in the unsmoothed data, except in rare
cases when it was necessary to suppress
%
``ringing'' (spectral
sidelobes) from a strong, spectrally unresolved feature.
%

Consistent with the primary motivation for this project, emphasis in
observing time was given to sources with strong absorption
uncontaminated by maser emission at either 6030 or 6035 MHz.
Secondary motivations included detecting sources with weaker
absorption and measuring Zeeman splitting of masing sources in massive
star-forming regions.  Additionally, one carbon star (IRC +10216) and
one Mira variable star (R Leonis) were observed.  A complete list of
sources observed with the aforementioned setup is provided in Table
\ref{source-table}.

\section{Results}

As noted in Table \ref{source-table}, some combination of emission or
absorption was detected toward most of the 29 sources we observed in
at least one of the main-line 6.0 GHz transitions.  Sixteen of our
sources had clearly detectable maser emission, and several of these
sources also had detectable broad emission.  Parameters of the
Gaussian fits to the emission features are given in Table
\ref{maser-table}.

Pairs of maser features that we identify as probable Zeeman splitting
are listed in Table \ref{zeeman-table} along with implied values of
the three-dimensional magnetic field strength.  Maser Zeeman splitting
also provides information on the line-of-sight direction of the
magnetic field: positive field values indicate that the magnetic field
points in the hemisphere away from the Sun and negative values toward
the Sun.  We identify Zeeman pairs in every source in which masers are
positively detected with the exception of G5.886$-$0.393, in which
the only two masers we detect are at the same velocity to within the
fit accuracy.

In order to check the consistency of our flux scale with previous
studies, we have compared our absorption parameters with those of
several other studies.  We find that our flux scale is lower than that
of \citet{gardner83a}, mixed though on average slightly lower than
\citet{guilloteau84}, and slightly higher than B97, all of which used
the Effelsberg 100 m telescope.  Our detected absorption for
G5.886$-$0.393 is consistent with the flux densities measured in ATCA
observations by \citet{caswell01}.  Variability precludes the reliable
use of maser fluxes as an independent check on flux calibration.

\setcounter{table}{2}
\begin{table}
\caption{Zeeman Pairs}
\label{zeeman-table}
\centering
\begin{tabular}{lrrr}
\hline\hline
 & Transition & $v_\mathrm{LSR}^\mathrm{a}$ & $B$ \\
Source & (MHz) & (km~s$^{-1}$) & (mG) \\
\hline\hline
W3(OH)$^\mathrm{b}$   & 6030 & $-$69.92 & $-$4.4 \\
                &      & $-$69.58 & $-$3.8 \\
Sgr B2M         & 6030 &    65.54 & $-$4.8 \\
                &      &    70.42 & $-$5.4 \\
                &      &    71.90 & $-$8.0$^\mathrm{c}$ \\
                & 6035 &    62.22 & $+$5.9 \\
                &      &    63.42 & $-$5.7 \\
                &      &    64.82 & $-$7.1 \\
                &      &    67.18 & $-$6.9 \\
                &      &    70.45 & $-$6.9 \\
                &      &    70.83 & $-$7.3 \\
                &      &    71.94 &$-$10.8 \\
M17             & 6030 &    21.52 & $+$0.8 \\
                & 6035 &    21.41 & $+$0.8 \\
G28.199$-$0.048 & 6030 &    94.70 & $+$7.0 \\
                &      &    96.12 & $-$1.4 \\
                & 6035 &    94.70 & $+$7.0 \\
                &      &    98.04 & $+$8.9 \\
W48             & 6035 &    42.95 & $-$0.7 \\
                &      &    43.30 & $-$0.9 \\
G45.122$+$0.133 & 6030 &    53.70 & $-$4.8 \\
                &      &    54.33 & $-$4.5 \\
                & 6035 &    53.70 & $-$4.4 \\
                &      &    54.27 & $-$4.0 \\
                &      &    55.45 & $-$1.2 \\
W51 e/d         & 6030 &    52.53 & $+$5.1 \\
                &      &    53.15 & $+$5.1 \\
                & 6035 &    52.80 & $+$6.7 \\
                &      &    55.33 & $+$4.6 \\
                &      &    55.81 & $+$9.1 \\
                &      &    57.59 & $+$5.5 \\
                &      &    63.11 & $+$2.3 \\
                &      &    64.04 & $+$4.0 \\
K3-50           & 6030 & $-$19.45 & $-$8.8 \\
                & 6035 & $-$19.53 & $-$8.7 \\
                &      & $-$18.80 & $-$6.9 \\
ON 1            & 6030 &     0.16 &$-$13.5 \\
                &      &    13.99 & $-$4.8 \\
                & 6035 &  $-$0.42 &$-$10.4$^\mathrm{c}$ \\
                &      &     0.12 &$-$12.8$^\mathrm{c}$ \\
                &      &     0.79 &$-$11.1$^\mathrm{c}$ \\
                &      &     1.53 & $-$5.3 \\
                &      &     1.96 & $-$5.2 \\
                &      &    13.91 & $-$3.8 \\
                &      &    14.49 & $-$0.8 \\
                &      &    15.30 & $-$5.0 \\
DR 20           & 6030 & $-$11.05 & $-$3.5 \\
                & 6035 & $-$11.10 & $-$3.0 \\
                &      &  $-$9.75 & $-$4.1 \\
W75 N           & 6035 &     7.85 & $+$7.1 \\
                &      &     9.54 & $+$4.6 \\
W75 S           & 6035 &  $-$2.31 & $-$2.3 \\
                &      &     3.33 & $-$4.8 \\
LDN 1084        & 6035 & $-$62.72 & $+$4.1 \\
                &      & $-$61.48 & $+$0.9 \\
NGC 7538        & 6035 & $-$56.88 & $+$0.9 \\
\hline
\end{tabular}
\begin{list}{}{}
\item[$^\mathrm{a}$] Systemic velocity ($(v_\mathrm{RCP} +
  v_\mathrm{LCP})/2$).
\item[$^\mathrm{b}$] The spectra are complicated at velocities between
  $-49$ and $-42$~km~s$^{-1}$, and we cannot identify individual
  Zeeman pairs with any certainty.  See Sect.\ \ref{sourcenotes}.
\item[$^\mathrm{c}$]Pairing ambiguity.
\end{list}
\end{table}

\begin{table}
\caption{Absorption Parameters}
\label{abs-table}
\centering
\begin{tabular}{lccrr}
\hline\hline
 & Transition & Absorption & $v_\mathrm{LSR}$ & $\Delta v$ \\
Source & (MHz) & (Jy) & (km s$^{-1}$) & (km s$^{-1}$) \\
\hline
W3 Cont         & 6030 & $-$0.03 & $-$43.83 & 4.14 \\
                &      & $-$0.29 & $-$39.58 & 2.64 \\
                &      & $-$0.11 & $-$37.86 & 2.59 \\
                & 6035 & $-$0.03 & $-$43.45 & 4.05 \\
                &      & $-$0.31 & $-$39.98 & 2.44 \\
                &      & $-$0.16 & $-$38.10 & 2.65 \\
G5.886$-$0.393  & 6030 & $-$0.10 & $-$19.76 & 6.51 \\
                &      & $-$0.12 &  $-$8.70 &25.52 \\
                & 6035 & $-$0.13 & $-$18.63 &10.49 \\
                &      & $-$0.10 &  $-$6.24 &25.33 \\
G10.624$-$0.385 & 6030 & $-$0.11 &  $-$6.67 & 5.19 \\
                & 6035 & \nodata$^\mathrm{a}$ &\nodata &\nodata \\
W48             & 6030 & $-$0.27 &    41.96 & 3.22 \\
                & 6035 & $-$0.26 &    42.08 & 3.05 \\
G45.122$+$0.133 & 6030 & $-$0.04 &    54.50 & 3.23 \\
                &      & $-$0.07 &    57.98 & 3.29 \\
                & 6035 & $-$0.08 &    58.17 & 3.61 \\
W51 e/d         & 6030 & $-$0.27 &    59.75 & 6.97 \\
                & 6035 & $-$0.22$^\mathrm{b}$ &    60.22 & 6.36 \\
K3-50           & 6030 & $-$0.12 & $-$25.21 & 7.84 \\
                &      & $-$0.08 & $-$24.60 & 4.00 \\
                & 6035 & $-$0.13 & $-$25.56 & 6.97 \\
                &      & $-$0.08 & $-$24.39 & 3.70 \\
DR 21           & 6016 & $-$0.06 & $-$10.63 & 8.90 \\
                &      & $-$0.13 &  $-$5.17 & 5.22 \\
                & 6030 & $-$0.05 & $-$13.61 & 2.25 \\
                &      & $-$0.44 & $-$10.25 & 9.91 \\
                &      & $-$0.77 &  $-$4.99 & 4.93 \\
                &      & $-$0.09 &  $-$1.98 & 2.10 \\
                & 6035 & $-$0.05 & $-$13.64 & 2.08 \\
                &      & $-$0.51 & $-$10.16 &10.27 \\
                &      & $-$0.79 &  $-$5.00 & 4.82 \\
                &      & $-$0.13 &  $-$2.26 & 2.08 \\
\hline
\end{tabular}
\begin{list}{}{}
\item[$^\mathrm{a}$]The low velocity wing of absorption is detected
   below $-5$~km~s$^{-1}$, but maser contamination and low
   signal-to-noise preclude fitting this feature.
\item[$^\mathrm{b}$]Region of absorption is heavily contaminated with
   emission, which may compromise accuracy of fit.
\end{list}
\end{table}

\begin{table}
\caption{Zeeman Splitting Deduced from OH Absorption}
\label{abs-zeeman-table}
\centering
\begin{tabular}{lrr}
\hline\hline
 & Transition & $B^\mathrm{a}$ \\
Source & (MHz) & (mG) \\
\hline
W3 Cont & 6030 & $< 0.5$ \\
        & 6035 & $< 0.6$ \\
DR 21   & 6030 & $< 0.3^\mathrm{b}$ \\
        & 6035 & $< 0.4$ \\
K3-50   & 6030 & $-1.1 \pm 0.3$ \\
        & 6035 & $< 0.9$ \\
\hline
\end{tabular}
\begin{list}{}{}
\item[$^\mathrm{a}$] Values are $3\,\sigma$ upper limits on $|B|$ for
     nondetections.
\item[$^\mathrm{b}$] Systematics may dominate.  See Sect.\
     \ref{zeeman-abs} for details.
\end{list}
\end{table}

\subsection{Absorption\label{absorption}}

Absorption features were detected toward eight sources, as listed in
Table \ref{abs-table}.  In some sources, such as W51 e/d at 6035 MHz,
the ability to fit absorption features is hampered by the presence of
maser features in the same velocity region as the absorption.  When
the absorption is sufficiently strong and relatively uncontaminated by
maser emission, normally two or even three Gaussian components are
required to fit the absorption.  This appears to be true at 13434 and
13441 MHz as well.  When absorption is detected with sufficient
signal-to-noise and spectral resolution, more than one Gaussian
fit component is required \citep{frm05}.

As mentioned in Sect.\ \ref{intro}, primary among our motivations for this
work was to obtain magnetic field estimates in the absorbing OH
material in high-mass star-forming regions.  To this end, we selected
three sources with strong absorption for deep integration: W3 Cont, DR
21, and K3-50.  These sources are uncontaminated by maser emission,
with the exception of the high-velocity wing of the absorption in K3-50.

The theory of Zeeman splitting in absorption is discussed in detail in
the literature \citep[e.g.,][]{troland82,sault90}.  The Stokes V
spectrum is related to the derivative of the Stokes I spectrum via the
equation
\[
V = -C \, \frac{dI}{d\nu} \, B_\parallel,
\]
where $V$ and $I$ are the Stokes V and I flux densities, $C$ is the
Zeeman splitting coefficient between the two $\sigma$-components, and
$B_\parallel$ is the strength of the line-of-sight component of the
magnetic field.  In principle, the Stokes V spectrum is simply half
the difference of the LCP and RCP spectra.  In practice, it is often
the case that the LCP and RCP spectra have different gains.  If this
gain difference is not corrected for, the Stokes V spectrum will be
dominated by a scaled version of the Stokes I spectrum.  For each
source and transition, we scaled one polarization to remove this
effect.  Note that this will not create a false detection.  The
derivative of Stokes I produces an ``S-curve,'' or adjacent positive
and negative spectral features, which is duplicated in the Stokes V
spectrum in the presence of a magnetic field.  However, an incorrect
scaling factor between the two circular polarizations will produce
either a positive or a negative feature in Stokes V, \emph{but not
both}.

For each source, the magnetic field was determined by determining the
best multiple-Gaussian fit to the Stokes I spectrum, computing the
derivative of the fit, and least-squares fitting the resulting
derivative to the Stokes V spectrum.  Results are shown for W3 Cont in
Figs.\ \ref{w3-6030-iv} and \ref{w3-6035-iv}, K3-50 in Figs.\
\ref{k350-6030-iv} and \ref{k350-6035-iv}, and for DR 21 in Figs.\
\ref{dr21-6030-iv} and \ref{dr21-6035-iv}.  The top panel of each plot
shows the Stokes I spectrum along with the best-fit Gaussian
parameters as listed in Table \ref{abs-table}.  The bottom panel of
each plot shows the Stokes V spectrum as well as the derivative of the
fits to the Stokes I spectrum.  This derivative is scaled to fit the
data for K3-50, in which a marginal positive result is obtained at
6030 MHz.  (In this case, the derivative scaling factor and the
RCP/LCP scaling factor were solved for simultaneously by least-squares
fitting.)  For the other sources the scaling is chosen in order to
most clearly show the functional form of the derivative.  Implied
magnetic fields and upper limits are quoted in Table
\ref{abs-zeeman-table}.

Up to four Gaussians were fit to each Stokes I absorption profile.
Since absorption is produced by an integrated column density of
molecules at different velocities, it is probable that Stokes I
absorption profiles deviate from sums of a small number of pure
Gaussians when observed with sufficient sensitivity.  But the
magnitude of the expected Stokes V profile due to Zeeman splitting is
only a few percent of Stokes I.  Thus, our magnetic field estimates
are not highly sensitive to minor errors in fitting Stokes I profiles.

\subsection{Source notes\label{sourcenotes}}

For sources in large complexes of \ion{H}{ii}~regions, our pointing
center is occasionally offset from the coordinates of the associated
source in order for our 130\arcsec\ beam to encompass several sources
or a larger area of continuum flux.  Exact pointing centers are listed
in Table \ref{source-table}.

\begin{figure}
\resizebox{\hsize}{!}{\includegraphics{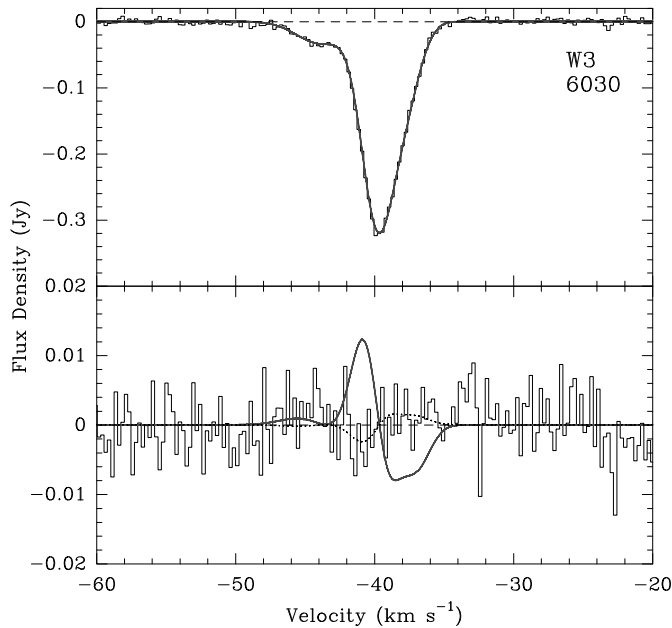}}
\caption{\textbf{Top:} Stokes I spectrum of W3 Cont at 6030 MHz.  The curve
  shows the best fit to the data, which have been Hanning weighted.
  Parameters of the fit are given in Table \ref{abs-table}.  \textbf{Bottom:}
  Stokes V spectrum of W3 Cont at 6030 MHz.  The curve shows the
  derivative of the Stokes I fit, scaled to a magnetic field of
  $+1.0$~mG in order to clearly show the functional form.  This curve
  can be scaled by an arbitrary multiplicative factor (including
  negative values); the scaling is proportional to the line-of-sight
  component of the magnetic field in the absorbing material.  The
  best fit scaling of $-0.2$~mG, indicated by the dotted line, is not
  statistically significant.
\label{w3-6030-iv}}
\end{figure}

\begin{figure}
\resizebox{\hsize}{!}{\includegraphics{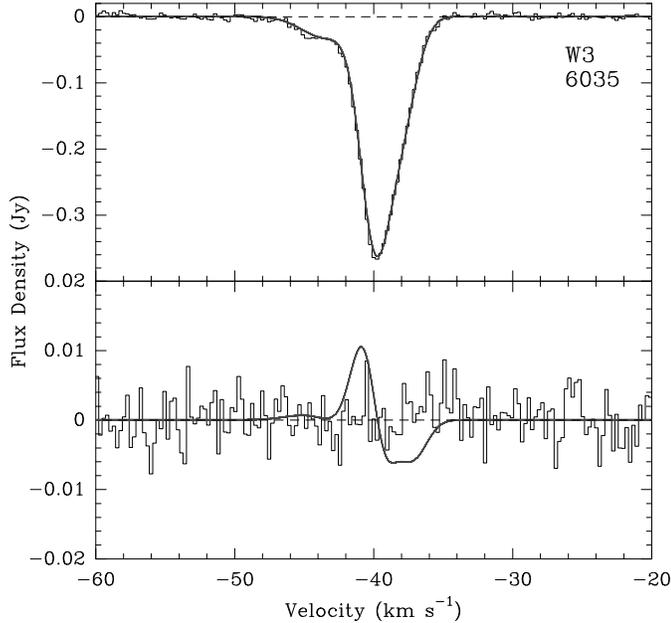}}
\caption{Stokes I and V spectra of W3 Cont at 6035 MHz.  See Fig.\
  \ref{w3-6030-iv} for details.
\label{w3-6035-iv}}
\end{figure}

W3 Cont -- The coordinates listed in Table \ref{source-table} are
approximate.  To observe this source, we peaked up on the continuum
nearest this position, located approximately 10\arcsec\ away, which
lies in the direction of component W3A in the labelling scheme of
\citet{harris76}.  Absorption spectra are shown in Figs.\
\ref{w3-6030-iv} and \ref{w3-6035-iv}.  We find that 3 Gaussian
components are required to fit the absorption.  The velocities of the
main absorption component in each transition agree generally with
those of \citet{gardner83a}, \citet{guilloteau84}, and B97.
Ground-state absorption at 1667 MHz is centered near $-38.5$~km~s$^{-1}$\
\citep{troland89}.  Weak 4750 MHz emission as well as 4765 MHz masers
are seen in the velocity range of absorption \citep{gardner83b}.  We
are able to obtain a $3\,\sigma$ upper limit of 0.5 mG on the
line-of-sight component of the magnetic field in the absorbing OH
material.

W3(OH) -- Substantial ringing exists at 6035 MHz due to strong,
narrow masers.  Ringing is so substantial in RCP that the data must be
Hanning weighted twice, degrading effective resolution to $0.49$~km~s$^{-1}$,
or much broader than a typical maser linewidth.  At this resolution,
it is difficult to identify individual maser components and futile to
attempt to identify Zeeman pairs.  Even at 6030 MHz, where ringing is
far less severe, overlap of maser components in the spectral domain
precludes accurate fitting of maser components as well as unambiguous
identification of Zeeman pairs.  We refer the reader to previous
interferometric work by \citet{moran78} and \citet{desmurs98} for more
precise maser and Zeeman pair identification.

\begin{figure}
\resizebox{\hsize}{!}{\includegraphics{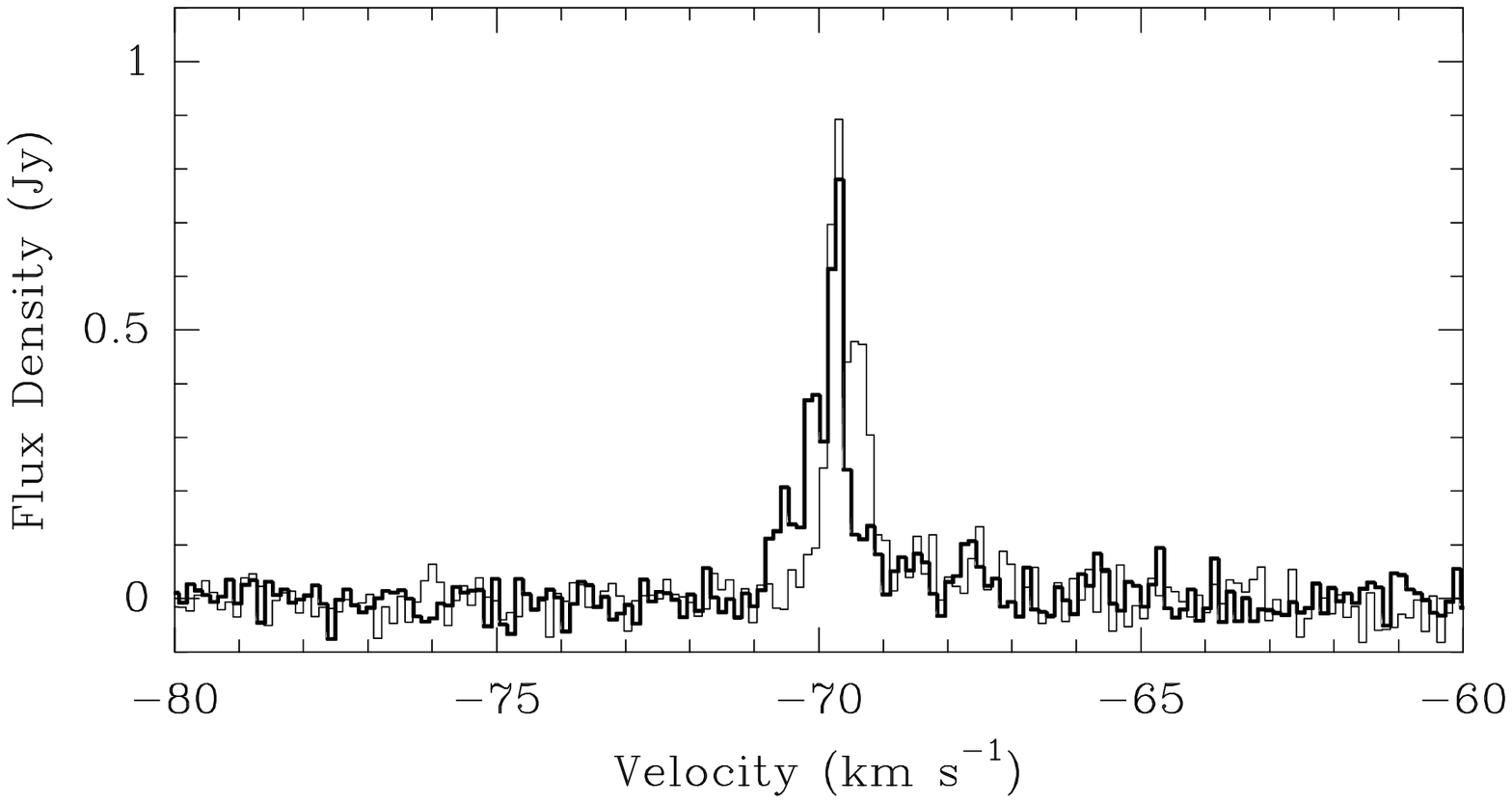}}
\caption{Spectra of new features of W3(OH) detected at 6030 MHz.  The
  data are not Hanning weighted.  Strong emission at higher velocity
  is not shown.  The bold histogram shows RCP data, and the normal
  histogram shows LCP data.  The RCP data are shifted to lower
  velocity than the LCP data, indicating a negative magnetic field.
\label{w3oh-figure}}
\end{figure}

We find new maser features near $-70$~km~s$^{-1}$\ in the 6030 MHz
transition, as shown in Fig.\ \ref{w3oh-figure}.  It is unlikely that
these masers are associated with the ultracompact \ion{H}{ii} region of
W3(OH).  The masers around W3(OH) have radial velocities of $-45 \pm
5$~km~s$^{-1}$ \citep[e.g.,][]{arm} and are observed to be expanding
\citep{bloemhof92} and rotating \citep{wright04} at several km~s$^{-1}$, not
tens of km~s$^{-1}$.  The association of these masers with W3(OH) proper
cannot be conclusively ruled out because former studies of W3(OH) have
not had the velocity coverage to have detected a feature at
$-70$~km~s$^{-1}$.  It is more plausible that these masers are associated
with the Turner-Welch (TW) object located $\approx 6$\arcsec\ east of
W3(OH) \citep{turner84}.  This source is also located well within our
130\arcsec\ beam, and the water masers span a velocity range of over
75~km~s$^{-1}$, including a feature at $-70$~km~s$^{-1}$\
\citep{cohen79}.  These 6030 MHz features are discussed in further
detail in Sect.\ \ref{maserstats}.

The detection of these OH masers at an unusual velocity prompted us to
obtain follow-up exploratory observations of all four ground-state
transitions of OH as well as the 4750 and 4765 MHz transitions with
the VLA\footnote{The National Radio Astronomy Observatory is a
facility of the National Science Foundation operated under cooperative
agreement by Associated Universities, Inc.}.  We observed the
ground-state transitions in 256 spectral channels of 3.052 kHz each
and the excited-state transitions in 128 channels of 6.104 kHz each,
giving us a velocity range of approximately 130 and 50~km~s$^{-1}$\ centered
at $-55$~km~s$^{-1}$\ for the ground-state and excited-state transitions,
respectively.  In no instance did we detect emission at $-70$~km~s$^{-1}$\ or
at any other velocity outside the systemic range of velocities from
approximately $-40$ to $-50$~km~s$^{-1}$.  Our $3\,\sigma$ noise detection
levels were approximately 60 to 75 mJy for all transitions.

Orion KL -- The data are suggestive of a weak ($\lesssim 0.5$~Jy)
maser at $+4.0$~km~s$^{-1}$\ in both polarizations of the 6035 MHz
transition, but the signal-to-noise ratio is too low to claim a
detection.  This velocity is consistent with the velocity of masers at
1665 MHz \citep[e.g.,][]{hansen77}.  B97 do not make note
of a detection in this source, although their 6035 MHz noise limits
for Orion KL are better than ours by nearly a factor of two.

R Leo -- No masers were detected in R Leo.  The bright ($> 1$ Jy)
ground-state masers in this source disappeared in 2002 and have only
reappeared as much weaker masers since \citep{lewis04}.

Sgr B2M -- Our spectra are qualitatively similar to those presented in
\citet{caswell95} and \citet{caswell03}, although ringing effectively
degrades our spectral resolution at 6035 MHz RCP.  \citet{caswell03}
identifies a Zeeman pair of $-5$~mG at $70.7$~km~s$^{-1}$\ in his 6030
MHz data and finds similar magnetic fields in the high-velocity 6035
MHz data.  All Zeeman pairs indicate a magnetic field pointing in the
hemisphere toward the Sun, with the exception of a single Zeeman pair
at $62.2$~km~s$^{-1}$, also noted by \citet{caswell03}.  ATCA
observations by \citet{caswell97} indicate that the 6035 MHz masers
trace two distinct \ion{H}{ii} regions: features below $\approx
68$~km~s$^{-1}$\ are associated with source G0.666$-$0.035
%
(the bright continuum near the center of Sgr B2M),
%
and features above this velocity with G0.666$-$0.029
%
(the weaker continuum to the west of Sgr B2M).
%

Because Sgr B2M is a source with many maser features, it is probable
that spatially-separated regions of maser emission at different
velocities are blended together in our beam.  Since we do not
spatially resolve the emission, we cannot always identify Zeeman pairs
%
unambiguously.
%
For instance, the marked Zeeman pair in Table
\ref{zeeman-table} consists of a feature at 72.21~km~s$^{-1}$\ in LCP
and 71.58~km~s$^{-1}$\ in RCP.  We choose this particular pairing over
the other possibilities (71.99 or 72.43~km~s$^{-1}$\ in RCP) in part
because the sign and magnitude of the magnetic field more closely
match those of the magnetic fields determined from other Zeeman pairs
near this velocity, both at 6030 and 6035 MHz.  But without
interferometric mapping, we cannot be certain that this pairing is
correct.

Sgr B2M was also observed for 36 min in a 20 MHz band centered near
5935 MHz.  This frequency range includes both the 5934.644
($^2\Pi_{3/2}, J = 5/2, F = 2 \rightarrow 2$) and 5938.967 MHz ($F = 3
\rightarrow 3$) main lines of $^{18}$OH as well as the 5938.901 MHz
($F, F^\prime = 5/2, 2 \rightarrow 3/2, 1$) line of $^{17}$OH.  The
dual-polarization noise in a Hanning-weighted 1.97~km~s$^{-1}$\ channel was 7
mJy.  The 5931.908 MHz ($F, F^\prime = 3/2, 2 \rightarrow 1/2, 1$)
line of $^{17}$OH is also in this range but is unobservable due to
extremely strong contamination from the H103$\alpha$ recombination
line at 5931.545 MHz.  No isotopomers of OH were detected.  This is as
expected, given a measurement of $261 \pm 20$ for the
[$^{16}$O]/[$^{18}$O] ratio in this source in the ground state of OH
\citep{whiteoak81}.  The expected strength of $^{18}$OH features in
Sgr B2M would be less than 1 mJy, assuming identical excitation
temperatures for $^{16}$OH and $^{18}$OH.  Since $^{17}$OH is even
less abundant than $^{18}$OH, the $^{17}$OH lines would be weaker still
\citep{valtz73,bujarrabal83}.

G5.886$-$0.393 -- Our spectra at both 6035 and 6030 MHz are similar
to the spectrum provided by \citet{caswell01}.  Our bandwidth spans an
effective velocity range of nearly 250~km~s$^{-1}$, which allows us to obtain
a good baseline even for the broad absorption features.  Our data are
consistent with the interpretation of the spectrum provided by Caswell.

G10.624$-$0.385 -- There are masers in both polarizations in the 6035
MHz transition in the velocity range from $-5$ to $+2$~km~s$^{-1}$.
Poor signal-to-noise and the overlap of multiple maser components,
especially in the upper half of this velocity range, prevent us from
being able to fit these features.  \citet{caswell95} and
\citet{caswell03} find weak 6035 MHz maser emission peaking at
$-0.7$~km~s$^{-1}$\ in Stokes I, a conclusion which our data support.
There is no clear Zeeman pattern, but if the peak of the RCP and LCP
emission are interpreted as a Zeeman pair, the implied magnetic field
is approximately $-4$~mG, consistent in direction with $-6.0$ and
%
$-2.86$~mG Zeeman pairs seen at 1667 MHz \citep{fish05,ruizvelasco06}.
%
We detect absorption in both transitions, although maser contamination
prevents a determination of fit parameters at 6035 MHz.  In the
$^2\Pi_{3/2}, J = 7/2$ \citep{matthews86,uchida01,frm05} and $J = 9/2$
\citep{walmsley86} transitions, absorption features are centered in
the velocity range $-2.5$ to $+1.0$~km~s$^{-1}$, where 6035 MHz maser
emission is strongest.

M17 -- The autocorrelator saturated during observations due to high
levels of continuum emission.  Flux densities derived for masers in
this source are of questionable accuracy, and we defer to the previous
data of B97.  As was sometimes seen in other scans with very high
continuum fluxes, the 6035 MHz RCP spectrum was flipped (i.e., the
continuum level was negative, and maser features were seen in
``absorption'').  This is clearly an instrumental artifact.  It is not
possible to obtain flux densities for these maser features, but the
velocities at which the features appear would not be expected to be
affected.  The velocities in both polarizations are consistent with
EVN maps of M17 \citep{desmurs98evn}.  Additionally, we identify one
Zeeman pair each at 6035 MHz and 6030 MHz at approximately the same
velocity and with equal splitting ($+0.8$~mG).  This too is consistent
with \citeauthor{desmurs98evn}, who find an upper limit of 1~mG for
the magnetic field.

G28.199$-$0.048 -- This is one of only two sources in our survey in
which we find a reversal of the line-of-sight direction of the
magnetic field.  \citet{caswell03} find a Zeeman pair with approximately
the same magnetic field and center velocity as the pairs we find at
$94.70$~km~s$^{-1}$.  They also find higher velocity features at 6030 MHz,
although B97 do not.

W48 (G35.200$-$1.736) -- At 6035 MHz, half of the velocity range of
absorption is contaminated by strong maser emission.  The RCP spectrum
is especially corrupted by ringing due to strong, narrow maser
features near $43$~km~s$^{-1}$.  B97 note similar problems in determining
absorption fit parameters.  Additionally, we detecte a weak 6035 MHz
maser in both polarizations coincident with the low-velocity wing of
the absorption.  Our spectra are similar to those of
\citet{caswell03}, although the Caswell 6035 MHz spectrum does not
include the weak maser features near 40.3~km~s$^{-1}$, whose LCP peak flux
density is half that of the 44.2~km~s$^{-1}$\ feature, clearly visible in the
Caswell spectrum.  Strong variability of this latter feature has been
noted by \citet{caswell01,caswell03}.  The 40.3~km~s$^{-1}$\ must have been
weaker by a factor of several in 2001 May and previous epochs.

G45.122$+$0.133 -- As with \citet{caswell03}, we find a consistent
Zeeman pattern at 6030 and 6035 MHz.  For most of the Zeeman pairs our
data suggest a stronger magnetic field ($\approx -4$~mG) compared to
the $-2.5$~mG determined by \citeauthor{caswell03}.  But precise
determination of the magnetic field strength, if indeed a single
magnetic field strength is responsible for the Zeeman pairs at $53.7$
and $54.3$~km~s$^{-1}$, is complicated by spectral blending of maser spots of
comparable flux density.  The velocity and width of the 6030 MHz
absorption agrees with a previous detection by \citet{guilloteau84} to
within their errors.  \citet{matthews86} and \citet{baudry02} also
detect absorption at 13434 and 13441 MHz in approximately the same
velocity range.

W51 e/d -- As with B97, our pointing center lies between
sources e and d.  We find eight Zeeman pairs each indicating a
magnetic field pointing in the hemisphere away from the Sun.  This is
consistent with B97 and \citet{desmurs98evn}, who each
find one Zeeman pair in 6035 MHz indicating a positive magnetic
field.  We identify Zeeman pairs at or near the velocities of these
features.  A pair of features at 6030 MHz in the B97
survey centered at $53.27$~km~s$^{-1}$\ with a splitting of $+5.1$~mG is
broadly consistent with a Zeeman pair of identical magnitude that we
detect at $53.15$~km~s$^{-1}$.  Absorption parameters are consistent with
those of \citet{guilloteau84} at 6030 MHz.  For comparison, the 13434
and 13441 MHz absorption is much narrower and redshifted by 3 to
4~km~s$^{-1}$\ \citep{baudry02}.

\begin{figure}
\resizebox{\hsize}{!}{\includegraphics{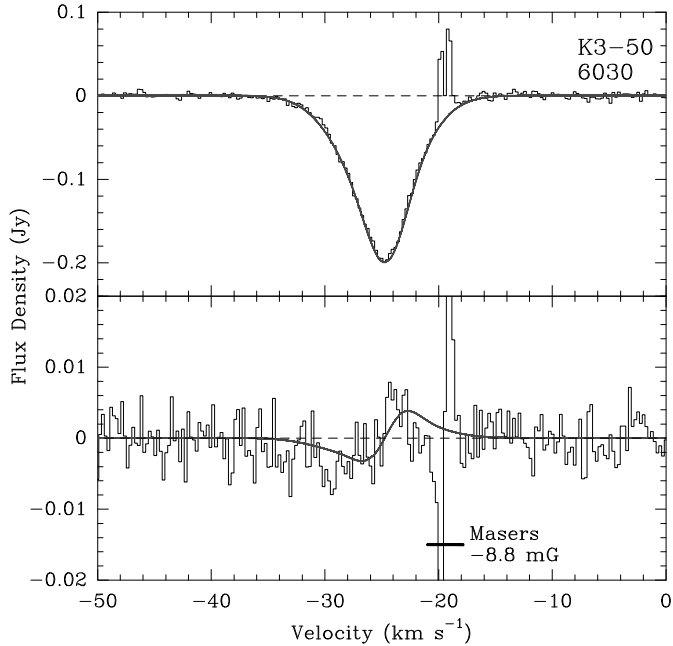}}
\caption{Stokes I and V spectra of K3-50 at 6030 MHz.  See Fig.\
  \ref{w3-6030-iv} for details.  The derivative curve in the bottom
  panel is scaled to $-1.1$~mG and represents a marginal ($<
  4\,\sigma$) detection.  The velocity range indicated by the dark bar
  was excluded from the fit due to maser contamination.  The magnetic
  field derived from Zeeman splitting of the masers is $-8.8$~mG.
\label{k350-6030-iv}}
\end{figure}

\begin{figure}
\resizebox{\hsize}{!}{\includegraphics{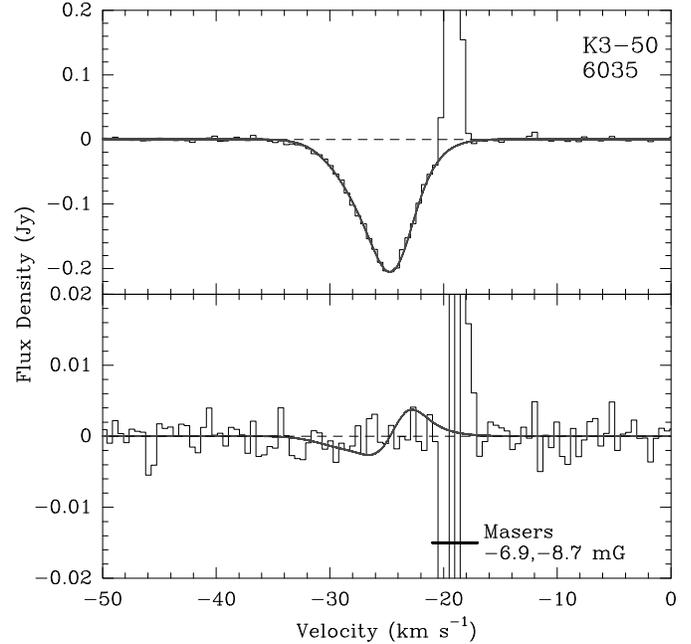}}
\caption{Stokes I and V spectra of K3-50 at 6035 MHz.  See Fig.\
  \ref{w3-6030-iv} for details.  The spectra have been Hanning
  weighted twice to eliminate ringing from strong maser components.
  The derivative curve in the bottom panel is scaled to $-1.1$~mG as
  determined from 6030 MHz (see Fig.\ \ref{k350-6030-iv}).  Unlike at
  6030 MHz, there is no detection.  The magnetic fields derived from
  Zeeman splitting of the masers are $-6.9$ and $-8.7$~mG.
\label{k350-6035-iv}}
\end{figure}

K3-50 -- Three Zeeman pairs are detected, all indicating a negative
magnetic field.  This is consistent with 1665 and 1667 MHz maser
mapping \citep{fish05}.  Previous observations have detected a single
absorption component at 6030 and 6035 MHz as well as in higher
transitions in the $^2\Pi_{3/2}$ ladder
\citep{guilloteau84,walmsley86,baudry02}.  We find that a single
Gaussian component is inadequate to fit our observed absorption,
although two components provide an excellent fit.  The 6030 and 6035
MHz spectra of \citeauthor{guilloteau84} did not have sufficient
signal-to-noise to permit the detection of the weaker component.  We
detect a magnetic field of $-1.1 \pm 0.3$~mG in the 6030 MHz
absorption (Fig.\ \ref{k350-6030-iv}), but we do not detect a
magnetic field in the 6035 MHz absorption (Fig.\ \ref{k350-6035-iv}).

ON 1 -- All Zeeman pairs in ON 1 indicate a negative
magnetic field, consistent with results in ground-state transitions
\citep[e.g.,][]{fish05}.  Consistent with prior 6030 and 6035 MHz
(B97) and 13441 MHz \citep{baudry02,frm05} observations,
most masers appear near two systemic velocities: 0 and $+14$~km~s$^{-1}$.
Additionally, we detect weak masers at intermediate velocities (5.5
and 7.7~km~s$^{-1}$) at 6035 MHz.

DR 20 (G80.864$+$0.421) -- Moderate spectral ringing is seen in 6035
MHz.  It is possible that there is another weak feature in both RCP
and LCP at slightly higher velocity than the strongest maser feature,
but there is insufficient resolution to claim detection after Hanning
weighting.  The masers identified by B97 are in excellent
agreement with this work.

W75 N -- As B97, we find a Zeeman pair with a magnetic
field splitting of slightly over $+7$~mG.  We also detect a Zeeman
pair centered near $9.5$~km~s$^{-1}$\ implying a field splitting of
$+4.6$~mG.  Interferometric maps at 1665 and 1667 MHz show masers
aligned along two axes, with negative magnetic fields in the east-west
axis and positive fields in the north-south axis
\citep[e.g.,][]{fish05}.  The magnetic fields measured at 6035 MHz
suggest that the excited-state masers appear predominantly in the
latter structure, probably associated with continuum source VLA 1.

W75 S -- We detect several features near $3$~km~s$^{-1}$, including a Zeeman
pair implying a magnetic field of $-4.8$~mG, not previously detected
by B97.  In the 1665 and 1667 MHz transitions, there is a
reversal of the line-of-sight direction of the magnetic field across
the source, with positive magnetic fields to the west and negative
magnetic fields to the east of the weak continuum source
\citep{fish05}.  This suggests that the Zeeman pairs, which imply
negative magnetic fields, may also be located to the east.  We do not
detect the possible feature they see at $-4.39$~km~s$^{-1}$\ in 6030 MHz RCP.

\begin{figure}
\resizebox{\hsize}{!}{\includegraphics{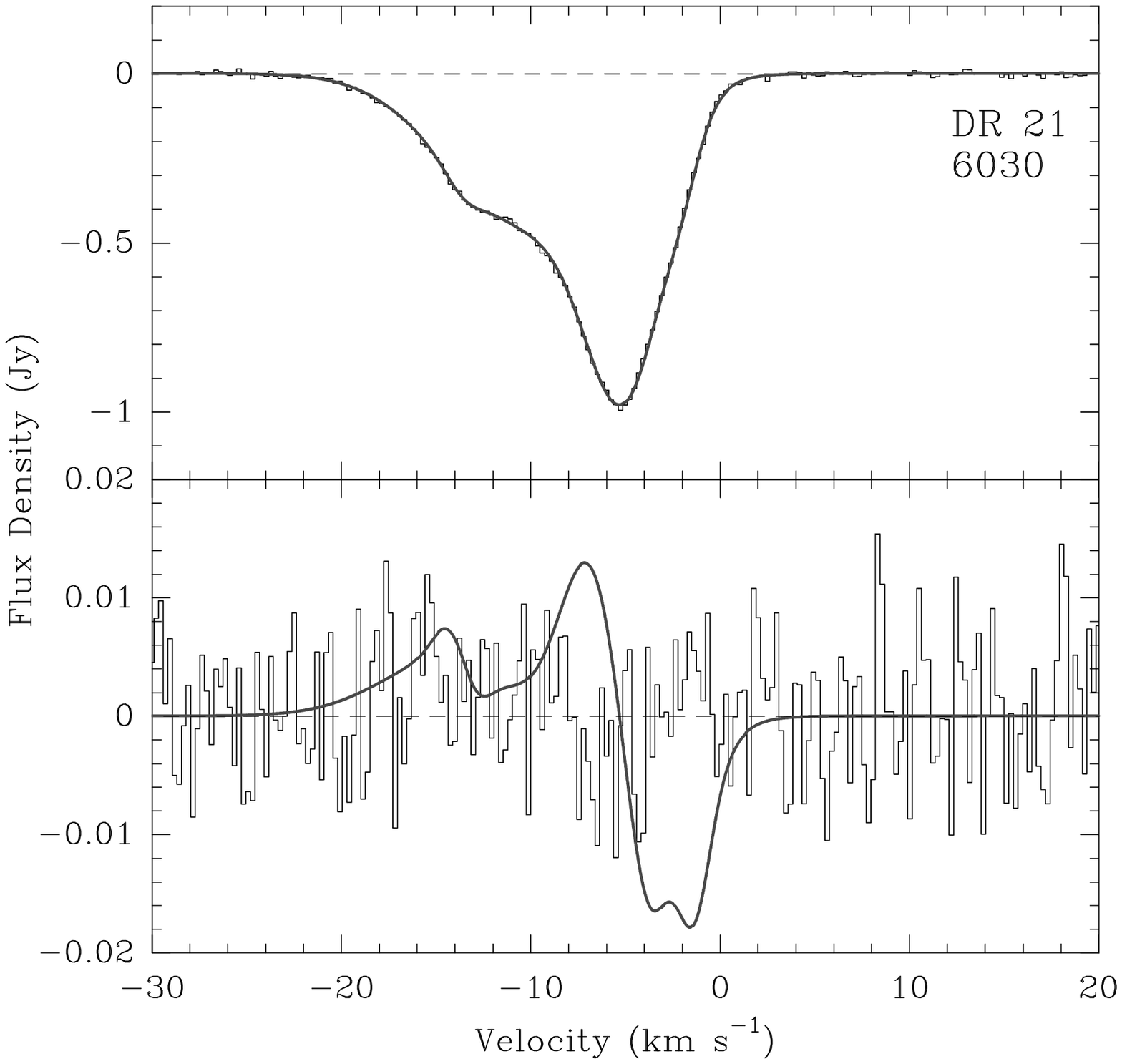}}
\caption{Stokes I and V spectra of DR 21 at 6030 MHz.  See Fig.\
  \ref{w3-6030-iv} for details.  The derivative curve in the bottom
  panel is scaled to $+1.0$~mG.
\label{dr21-6030-iv}}
\end{figure}

\begin{figure}
\resizebox{\hsize}{!}{\includegraphics{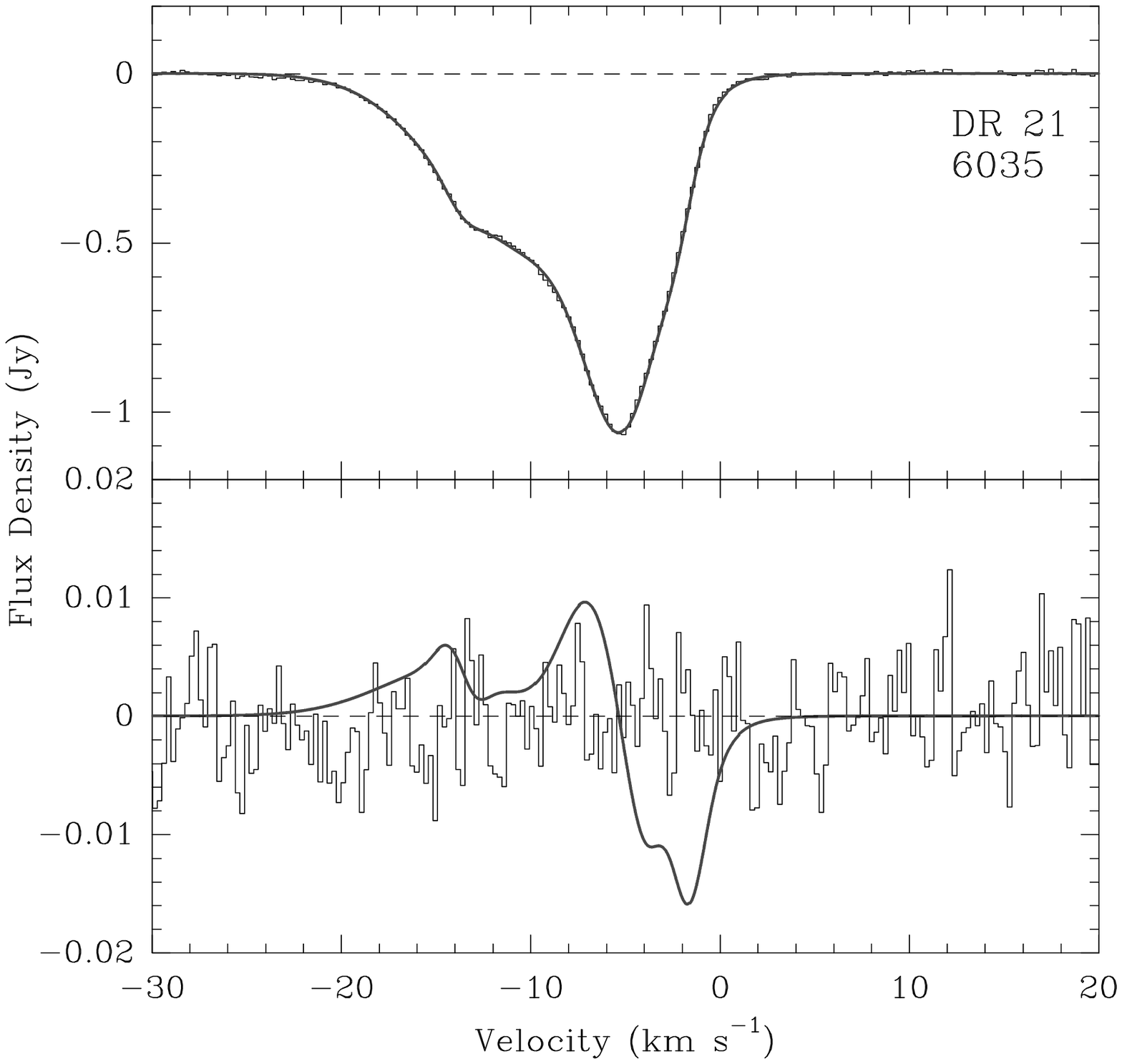}}
\caption{Stokes I and V spectra of DR 21 at 6035 MHz.  See Fig.\
  \ref{w3-6030-iv} for details.  The derivative curve in the bottom
  panel is scaled to $+1.0$~mG.
\label{dr21-6035-iv}}
\end{figure}

DR 21 -- The strongest absorption components we detect are
qualitatively consistent with previous results by
\citet{guilloteau84}, who also find two absorption components in each
of the main-line transitions: a feature near $-5$~km~s$^{-1}$\ and a slightly
broader feature near $-12$~km~s$^{-1}$\ that is weaker by a factor of two.
Spectra of Stokes I and V for the 6030 and 6035 MHz transitions are
shown in Figs.\ \ref{dr21-6030-iv} and \ref{dr21-6035-iv},
respectively.  \citet{matthews86} find absorption at 13434 and 13441
MHz at the velocity of the stronger absorption component, while
\citet{walmsley86} see 23826 MHz absorption at $-6.5$~km~s$^{-1}$.  At 6016
MHz, \citet{gardner83a} also see two absorption components.

LDN 1084 -- We identify two Zeeman pairs indicating magnetic fields of
$+0.9$ and $+4.1$~mG in the 6035 MHz transition.  B97
detect two 6035 MHz masers in opposite polarizations.  Interpreted as a
Zeeman pair, they would imply a magnetic field of $+1.4$~mG centered
at $-61.33$~km~s$^{-1}$.  The sign of this field would be consistent with
our findings.

S140 -- B97 detect a Zeeman pair of masers of flux density 0.4 Jy and
width 0.20~km~s$^{-1}$\ at 6035 MHz.  Neither these features nor any others
are present in our spectra.  Our $1\,\sigma$ noise level in a
Hanning-weighted 0.243~km~s$^{-1}$\ channel is 13 mJy.

Cep A -- We detect no maser emission in either transition to a $1\,
\sigma$ limit of 34 mJy in a Hanning-weighted 0.243~km~s$^{-1}$\ channel.
B97 detect multiple maser lines with flux densities of 0.1 to 0.6 Jy,
including two Zeeman pairs in the 6035 MHz transition.

NGC 7538 -- Our identification of a $+0.9$~mG Zeeman pair is
consistent with observations at 1665 MHz, in which a single Zeeman
pair of $+0.7$~mG is detected \citep{fish05}.  However, previous
observations have found Zeeman pairs with the opposite sense of
polarization \citep[e.g.,][]{lo75}.  In ground-state transitions,
\citet{hutawarakorn03} identify two Zeeman pairs of $-1.7$ and
$-2.0$~mG in source IRS 1, the pointing center of observations.  They
also find three Zeeman pairs with a line-of-sight magnetic field
reversal in IRS 11, located approximately half a beamwidth away from
our pointing center.

\section{Discussion\label{discussion}}

\subsection{6030 and 6035 MHz maser statistics\label{maserstats}}

We detect 6035 MHz masers in 16 sources and 6030 MHz masers in 11 of
these.  In all cases, the 6035 MHz masers are stronger than the 6030
MHz masers, whether measured in terms of the strongest maser component
or as an integrated flux of all maser spots in a transition.  In
nearly all cases when a 6030 MHz maser is observed, a stronger 6035
MHz maser is seen at approximately the same velocity, consistent with
B97 and observations by \citet{caswell03} of southern hemisphere
sources.  Diverse theoretical models agree that the set of parameter
space conducive to 6030 MHz maser action is almost wholly a subset of
that conducive to 6035 MHz maser action and that 6030 MHz masers
should almost always be weaker than their 6035 MHz counterparts
\citep[e.g.,][]{gray92,pavlakis00,cragg02}. 

However, the 6030 MHz features near $-70$~km~s$^{-1}$\ in W3(OH) have no
counterpart at 6035 MHz.  Theoretical models by \citet{pavlakis00}
suggest that the presence of 6030 MHz without corresponding 6035 MHz
maser emission is indicative of a velocity gradient greater than
1~km~s$^{-1}$.  This suggests that the low-velocity 6030 MHz masers may not
be associated with the main ultracompact \ion{H}{ii}\ region in W3(OH),
for which the turbulence is estimated to have a FHWM of 0.8~km~s$^{-1}$\ from
the spatial and velocity distribution of ground-state OH masers
\citep{reid80}.  For comparison, the turbulent component of the
velocity field of the water masers in the bipolar outflow from the TW
object is $\lesssim 10$~km~s$^{-1}$\ \citep{alcolea93}.
\citeauthor{pavlakis00} also predict that 1665 MHz masers should be
seen in regions containing 6030 but not 6035 MHz masers.  But no
ground-state maser emission in any transition is seen at $-70$~km~s$^{-1}$.
It is possible that current maser models are insufficient to explain
the appearance of masers at 6030 MHz with no corresponding features in
another OH transition.

\subsection{Zeeman pairs\label{zeemandiscussion}}

As with any single-dish study of OH masers, one must be careful
interpreting spectral offsets as Zeeman pairs.  In the 1665 and 1667
MHz transitions, the Zeeman splitting is large (0.590 and 0.354
km~s$^{-1}$~mG$^{-1}$ respectively), and a typical magnetic field of several
milligauss will split the two $\sigma$-components by many linewidths.
Velocity coherence of the amplifying material usually favors one
$\sigma$-component over the other, resulting in Zeeman pairs of very
unequal amplitude or the detection of only a single $\sigma$-component
\citep{cook66}.  Since high-mass star-forming regions with OH masers
frequently have many masers, including multiple masers at the same
velocity, single-dish efforts at identifying maser pairs are generally
inadequate.  In practice, even connected-element interferometry
provides insufficient spatial resolution to identify Zeeman pairs in
the ground-state main-line transitions (e.g., compare sources in
common between \citealt{fish03} and \citealt{fish05}).  At 6030 and
6035 MHz, the smaller Zeeman coefficients (0.0790 and 0.0564
km~s$^{-1}$~mG$^{-1}$ respectively) result in $\sigma$-components that are
usually separated by less than 0.8~km~s$^{-1}$, the turbulent dispersion
inferred from maser clusters for W3(OH) \citep{reid80}.  Thus, it is
expected that amplification of both $\sigma$-components of a Zeeman
pair occurs with similar gain, a hypothesis supported by the more
equal ratio of flux densities in the components of a 6.0 GHz Zeeman
pair as compared with ground-state Zeeman pairs.

On the other hand, the dearth of sources observed at 6.0 GHz at
milliarcsecond resolution make it difficult to determine whether this
hypothesis is justified.  To date, only four published sources have
been mapped at 6030 and/or 6035 MHz: W3(OH) \citep{desmurs98}, M17, ON
1, and W51 \citep[][for the latter three]{desmurs98evn}.  In W3(OH),
multiple strong, spatially-separated maser features in the same
polarization and transition were often seen at the same velocity.
\citeauthor{desmurs98} find 37 Zeeman pairs and several unpaired maser
components at 6030 and 6035 MHz between $-47.7$ and $-41.5$~km~s$^{-1}$.  The
ratio of fluxes between the two $\sigma$-components rarely exceeds two
and never exceeds three.  Magnetic fields thus measured range from
$+0.9$ to $+14.6$~mG.  Clearly, single-dish observations of W3(OH) are
insufficient in making sense of such rich maser fields.  In the other
three sources, \citeauthor{desmurs98evn} find far fewer maser spots
and Zeeman pairs than we do due to a high detection limit (0.4 Jy
beam$^{-1}$).  Even in ON 1, in which only four 6035 MHz RCP masers are
detected with the EVN, there is evidence that spatially-separated
maser features overlap in velocity.

From this, it seems reasonable to conclude that single-dish Zeeman
pairing at 6030 and 6035 MHz is trustworthy in spectrally
uncomplicated fields with comparable flux densities in opposite
circular polarizations, such as LDN 1084.  It is probable that sources
with numerous overlapping maser components, such as Sgr B2M and ON 1,
have many more Zeeman pairs than identified here.  Additionally,
ambiguities may cause misidentification of which maser components
constitute a Zeeman pair.  Judgment on the reliability of Zeeman pairs
in cases intermediate to these extremes awaits confirmation from
interferometric mapping.

\subsection{Zeeman splitting in absorption\label{zeeman-abs}}

\subsubsection{K3-50}

K3-50 is the only source in which we obtain a marginal positive
detection of a magnetic field in absorption.  We measure the magnetic
field to be $-1.1 \pm 0.3$~mG in absorption in the 6030 MHz
transition.  At 6035 MHz we obtain an upper limit ($3\,\sigma$) of
0.9~mG.  The previous upper limit on the magnetic field strength in
absorption was 2.9~mG from 13441 MHz observations \citep{frm05}.

Given the low signal-to-noise ratio of the 6030 MHz detection and the
lack of a detection at 6035 MHz, it is possible that the positive
detection at 6030 MHz is due to a systematic error.  One possible
source of error is an incorrect spectral baseline in one or both
circular polarizations.  As mentioned in Sect.\ \ref{observations},
spurious instrumental effects mimicking absorption were occasionally
seen in the 6030 MHz LCP spectra; while such cases were usually easily
identifiable, it is possible that the some LCP scans are contaminated
by similar instrumental effects too weak to be detected in a single
6-minute scan.

VLBA maps of the 1665 and 1667 MHz OH masers in K3-50 find six Zeeman
pairs: five implying full three-dimensional field strengths of $-2.6$
to $-2.9$~mG to the north and east of the \ion{H}{ii} region, and one
implying a magnetic field of $-7.5$~mG projected atop the northern
portion of the \ion{H}{ii} region \citep{fish05}.  In the 6030 and 6035
MHz transitions, Zeeman pairs indicate magnetic fields of $-5.3$ to
$-9.1$~mG (B97 and this work).  Thus, all OH maser data point to a
negative magnetic field in the region sampled by the masers.  However,
in the absence of an interferometric map of the positions of the 6.0
GHz masers, only one Zeeman pair is known to be located in projection
atop the \ion{H}{ii} region.  Without magnetic field measurements in
the western and southern portions of the source, it is unknown whether
there is a reversal of the line-of-sight direction of the magnetic
field across the source.  If there is, the magnetic field
strength measured in OH absorption would be an average value and would
therefore underestimate the actual line-of-sight magnetic field
strength.  Therefore, it is difficult to compare directly the magnetic
field strength measured in OH absorption with the field strength
derived from maser Zeeman splitting.

It is important to note that the interpretation of the Stokes V flux
in K3-50 as due to Zeeman splitting is predicated upon the assumption
that the OH absorption is thermal and unpolarized.
\citet{verschuur96} finds evidence that some sources may have weak,
nonthermal, partially polarized absorption components which could lead
to the erroneous estimates of magnetic field strength.  A more
definitive detection of Stokes V, as well as a sensitive
interferometric map of 6030 MHz absorption, may be necessary to
convincingly link our Stokes V detection to a magnetic field estimate.

\subsubsection{W3 Cont and DR 21}

We obtain upper limits for the line-of-sight component of the magnetic
field in two sources: W3 Cont and DR 21.  In each case, the
observations are consistent with an upper limit on the coherent
line-of-sight magnetic field strength of about half a milligauss.
Note that the upper limits assume that the magnetic field is constant
across the source.  This may be an incorrect assumption; the different
Gaussian absorption components likely correspond to physically
distinct clouds of material, each of which may have a different
average magnetic field strength and orientation.  Nevertheless, it
does not seem warranted to attempt to fit derivatives of individual
Gaussian absorption components scaled to different magnetic fields
when the measurement of Stokes V is consistent with zero, as is the
case for most of our spectra.

While the formal $3 \, \sigma$ upper limit on the line-of-sight
magnetic field strength is 0.3~mG in the 6030 MHz transition in DR 21,
the noise is likely dominated by systematic effects.  Unlike in W3
Cont or K3-50, or indeed even in the 6035 MHz transition in DR 21, it
was not possible to find a scaling constant such that Stokes V was
flat or proportional to the derivative of Stokes I in the velocity
region of absorption.  This could be due to two effects.  First, the
total absorbing region is fairly wide ($\approx 20$~km~s$^{-1}$), so it is
possible that the baselines of the LCP and RCP spectra are not
determined to high enough accuracy in the region of interest.  Second,
as noted in Sect.\ \ref{observations}, low-level artifacts resembling
absorption features occasionally appeared at 6030 MHz RCP.  Because
the actual absorption in DR 21 is so strong ($\approx 1$~Jy), the
presence of spurious instrumental absorption features at coincident
velocity may have escaped our detection.  In fact, nearly one-third of
the 6030 MHz DR 21 scans (representing all scans on source on one day)
were discarded because the LCP and RCP absorption shapes were
qualitatively different near $-15$~km~s$^{-1}$.

Zeeman splitting is detected in \ion{H}{i} absorption in several
different subregions of DR 21, implying an average magnetic field of
approximately $-0.4$~mG \citep{roberts97}.  The velocity range of the
Zeeman-split \ion{H}{i} features is generally coincident with the
velocity range in which excited-state OH absorption is detected.  They
detect three components in \ion{H}{i}: a broad negative wing between
$-20$ and $-9$~km~s$^{-1}$, roughly corresponding to the broad absorption
component we detect; a narrow feature near $-5$~km~s$^{-1}$\ with a FWHM of
$\approx 5$~km~s$^{-1}$, corresponding to an absorption component we detect
with similar parameters; and a narrow absorption component near
$+10$~km~s$^{-1}$, which we do not detect at all.  The magnetic field is most
strongly detected in \ion{H}{i} absorption in the outflow in the range
of velocities between $-20$ and $-10$~km~s$^{-1}$.  The number density in the
outflow is probably between $10^4$ and $10^5$~cm$^{-3}$
\citep[e.g.,][]{garden91}.  For comparison, the number density in the
OH absorption region exceeds $10^7$~cm$^{-3}$ \citep{jones94}.  For
any reasonable scaling of $B$ with $n^\kappa$, the magnetic field
should be detectable in OH absorption.

Why then do we detect no magnetic field on a submilligauss level
toward DR 21?  One possibility is that multiple sources are blended
together within our beam.  DR 21 is composed of several condensations
\citep[e.g.,][]{harris73}, although the most of the continuum flux
comes from regions A, B, and C, which are near the head of one of the
cometary \ion{H}{ii}\ regions \citep[see Fig.\ 1 of][]{cyganowski03}.  It is
possible that there exists one or more reversals of the line-of-sight
direction of the magnetic field across DR 21 which are being blended
together in our beam.  Our beam is much larger than the 10\arcsec\
effective resolution of \citeauthor{roberts97}, and it is possible
that areas of different magnetic field strength are blended together
in our beam.  However, \citeauthor{roberts97} obtain positive
detections of the magnetic field strength at several locations within
DR 21 and find that the magnetic field is everywhere consistent with
being $-0.4$~mG.  

As for W3 Cont, the largest negative and positive magnetic field
strengths measured in \ion{H}{i} are $-120$~$\mu$G and $+220$~$\mu$G
between W3 A and W3 B \citep{roberts93}, measured in the $-38$~km~s$^{-1}$\
velocity component.  \citet{crutcher99} also reports an \ion{H}{i}
measurement of 400~$\mu$G, which is below our detectability limit.  We
detect strong OH absorption near this velocity and almost no
absorption at $-46$~km~s$^{-1}$, the other velocity for which
\citeauthor{roberts93}\ obtain \ion{H}{i} Zeeman measurements.  Toward
the strongest continuum source (W3 A), they obtain a magnetic field
estimate of $-47$~$\mu$G.  All of these values are well below our
detection threshold, although the excited OH traces regions of higher
density and therefore likely higher magnetic field.  Nevertheless, the
line-of-sight magnetic field reversal between different sources in our
beam would serve to diminish the signal in Stokes V.

There are no OH maser measurements of the magnetic field in W3 Cont or
DR 21.  No masers have been detected in DR 21.  Weak 1665 MHz masers
have been detected in the W3 source G133.715$+$1.215 \citep{arm}.  If
interpreted as a Zeeman pair, the masers near the origin in the
\citeauthor{arm} map indicate a magnetic field strength of $B \approx
0.2$~mG, which is below our detection threshold.  For neither of these
sources then can we comment on the clumpiness of the material through
comparison of $B_\parallel$ with $B$.

\subsection{Are masers overdense?}

In general, OH masers exist in environments of higher magnetic field
strengths and have narrower line widths than OH absorption.  Thus,
in practice, OH maser Zeeman splitting is sensitive to the \emph{full,
three-dimensional} magnetic field strength, while absorption Zeeman
splitting is only sensitive to the \emph{line-of-sight component}
thereof.  Theoretically, the inclination of the magnetic field to the
line of sight at an OH maser site can be estimated from the linear
polarization fraction of the maser radiation \citep{goldreich73}.  The
ground-state OH masers that are projected atop the \ion{H}{ii} region
in K3-50 have no detected linear polarization, na\"{\i}vely suggesting
that the local magnetic field lies entirely in the plane of the sky
(i.e., that the line-of-sight component of the magnetic field is
zero).  However, the large percentage of maser spots with no
detectable linear polarization \citep[e.g.,][]{fish05} suggests that
OH masers preferentially amplify circularly-polarized radiation, a
result supported by diverse theoretical simulations
\citep[e.g.,][]{nedoluha90,gray94,elitzur96}.  In this case, magnetic
field inclinations derived from linear polarization fractions may be
misleading.  In fact, interpretation of the linear polarization from
ground-state OH masers may be impossible in some sources
\citep{fish06}.

There are now three sources with magnetic field estimates or upper
limits from OH absorption that are lower than the magnetic field
determined from OH maser Zeeman splitting: K3-50, W3(OH), and
G10.624$-$0.385 \citep{gusten94,uchida01,frm05}.  In each case the
line-of-sight field strength ($B_\parallel$) is significantly less
than $1/\sqrt{3}$ of the three-dimensional field strength ($B$)
derived from masers projected atop the continuum emission; this is the
value that would be expected if the magnetic field in these regions
were uniform and randomly oriented.  If magnetic beaming is important
for preferential amplification of $\sigma$-components of OH masers
\citep{gray94}, it is likely that $B_\parallel \approx B$ at maser
sites.

Assuming that an increased magnetic field strength indicates increased
density, it is likely that the density at OH maser sites is higher
than that of the surrounding material.  This is consistent with
observations of ammonia in massive star-forming regions.  Ammonia
absorption in G10.6$-$0.4 shows two components: a fairly uniform
component of low column density and a clumpy component of higher
column density \citep{sollins05}.  Observations of W3(OH) suggest that
ammonia and OH are cospatially distributed in a clumpy molecular
envelope \citep{reid87}.  Their observations support an
interclump-to-clump density ratio of $\sim 0.5$, which would imply an
interclump-to-clump magnetic field strength ratio of $\sim 0.7$
assuming that $B$ scales as $n^{0.5}$ \citep{crutcher99}.  Combining
with the projection effect, we would predict that $B_\parallel$ as
measured in the absorbing material would be between 0.4 and 0.7 times
$B$ measured from OH masers.  This is consistent with results for
W3(OH) \citep{gusten94,frm05}.  $B_\parallel$ as measured in
absorption appears to be even less than 0.4 times $B$ as measured from
Zeeman splitting in G10.624$-$0.385 \citep{frm05} and K3-50 (this
work), but the magnetic field in front of the UC\ion{H}{ii} region is
poorly sampled by OH masers in both of these sources.  Single-dish
absorption studies are sensitive only to the average value of
$B_\parallel$ in the absorbing material, weighted by OH density.  If
the magnetic field strength is variable or if there is a reversal of
the line-of-sight direction of the field across the source, absorption
measurements of $B_\parallel$ may be significantly smaller than $B$
derived from a single maser Zeeman pair.

Theoretical models support our conclusion that the single most
important physical factor discriminating between OH absorption and
maser emission is the density of the material.  \citet{cesaroni91}
find that OH can undergo a sharp change from strong absorption to
strong maser emission with a small change in density.  For most
transitions the critical density is between $10^5$ and
$10^6$~cm$^{-3}$ and the excitation temperature changes from
$T_\mathrm{ex} \gtrsim 0$ (efficient absorption) to $T_\mathrm{ex}
\lesssim 0$ (near maximum inversion) when the density
($n_{\mathrm{H}_2}$) changes by less than a factor of two across this
value.  \citeauthor{cesaroni91} find that thermalization occurs at
$n_{\mathrm{H}_2} \approx $ several $\times 10^7$~cm$^{-3}$, and OH
will be seen in absorption above this density.  Their results are
sensitive to the presence of dust; for instance, $T_\mathrm{ex}$ is
negative for two disjoint ranges of density for the 1667 MHz
transition.  \citet{pavlakis96,pavlakis00} explore a larger range of
physical conditions and find qualitatively similar maser behavior.

While observations of K3-50, W3(OH), and G10.624$-$0.385 argue
strongly for masers being overdense as compared to the OH seen in
absorption, the picture in DR 21 is less clear.  The best-fit density
($n_{\mathrm{H}_2} = 1.1$ -- $2.5 \times 10^7$~cm$^{-3}$) derived by
\citet{jones94} for DR 21 is difficult to reconcile with the low upper
limit on the magnetic field strength that we calculate from OH
($|B_\parallel| < 0.4$~mG) and that \citet{roberts97} calculate from
\ion{H}{i} ($|B_\parallel| = 0.4$~mG).  If the magnetic field strength
scales as $B \propto n^\kappa, \kappa \approx 0.5$ \citep{crutcher99},
we would expect a magnetic field strength at least ten times greater
than this value.  Furthermore, it is unclear why OH at this density
would be seen in absorption.  The models of \citet{cesaroni91}, which
assume a ratio of OH/H$_2$ consistent with the \citet{jones94} results
and a dust temperature only marginally higher than
\citeauthor{jones94}, indicate that many lines of OH, including the
6035 MHz transition, should be inverted in this density range.  In
contrast, only the 4.7 GHz ($^2\Pi_{1/2}, J = 1/2$) lines are seen in
weak, broad emission \citep{gardner83b}, while all other detected
transitions are seen in absorption \citep[see][]{jones94}.  The large
\emph{compact} (as opposed to \emph{ultracompact}) \ion{H}{ii} regions
in DR 21 argue for a density that is too low to sustain OH maser
emission or detectable OH absorption Zeeman splitting.

The significance of the density contrast between masers and the
ambient medium cannot be overstated.  Since masers are overdense
clumps of material, their observed proper motions can reliably be
interpreted as material motions.  Maser proper motions have been an
important tool both in measuring material motions in sources
\citep[e.g.,][]{bloemhof92} as well as for obtaining geometric
parallaxes independent of Galactic rotation curves \citep{hachisuka06}
and the cosmic distance scale \citep{herrnstein99,brunthaler05}.  Were
masers the result of chance lines of coherence due to random velocity
fluctuations in an otherwise homogeneous medium \citep{deguchi82},
observed proper motions could very well be unrelated to physical
motions of material.  Our results suggest that masers occur in
higher-density clumps of material.  While we cannot rule out the
possibility of beam averaging of opposite senses of the magnetic field
in some sources, our results are consistent with the interpretation of
the magnetic field derived from OH absorption in W3(OH)
\citep{gusten94,frm05}, a source whose magnetic field is well sampled
by OH masers \citep{wright04b}.  Combined with the observed
persistence of maser spot shapes between epochs \citep{bloemhof96},
our results constitute strong evidence that maser proper motions do
indeed indicate real, material motions.

\section{Conclusions}

We have observed 29 sources, primarily high-mass star forming regions,
in order to detect 6030 and 6035 MHz OH absorption and emission.  We
find maser emission in 16 sources and absorption in 8.  When 6030 MHz
maser emission is seen, it is almost always accompanied by stronger
6035 MHz maser emission.  However, W3(OH) exhibits an unexplained weak
6030 MHz maser at $-70$~km~s$^{-1}$ that appears to be unaccompanied by
maser emission at 6035 MHz, any of the four ground-state transitions,
or two of the $^2\Pi_{1/2}, J = 1/2$ transitions.

We have taken significantly long observations of three
sources with strong absorption to place submilligauss upper limits
($3\,\sigma$) on the average line-of-sight magnetic field strength.
We are able to obtain a weak tentative detection of $-1.1 \pm 0.3$~mG
in the 6030 MHz transition of K3-50, although our $3\,\sigma$ upper
limit in the 6035 MHz transition is lower than this value.  Consistent
with observations in the 13.4 GHz main-line transitions of OH
\citep{frm05}, the line-of-sight magnetic field strengths measured in
OH absorption are significantly lower than would be expected from
random orientation of the field, given three-dimensional magnetic
field strengths measured in OH masers.  Since the magnetic field
strength is correlated with density, this provides strong evidence
that OH masers occur in regions of enhanced density.  We conclude that
maser proper motions are a reliable indicator of physical material
motions.

\begin{acknowledgements}
Based on observations with the 100 m telescope of the MPIfR
(Max-Planck-Institut f\"{u}r Radioastronomie) at Effelsberg.
\end{acknowledgements}

\Online
\setcounter{table}{1}
\begin{longtable}{lrrrrr}
\caption{Detected Emission and Maser Emission}
\label{maser-table} \\
\hline\hline
 & Frequency & & Flux Density & $v_\mathrm{LSR}$ & $\Delta v$ \\
Source & (MHz) & Polarization & (Jy) & (km s$^{-1}$) & (km s$^{-1}$) \\
\hline
\endfirsthead
\caption{continued.} \\
\hline\hline
 & Frequency & & Flux Density & $v_\mathrm{LSR}$ & $\Delta v$ \\
Source & (MHz) & Polarization & (Jy) & (km s$^{-1}$) & (km s$^{-1}$) \\
\hline
 \endhead
 \hline
 \endfoot
W3(OH)          & 6030 & LCP & 0.70 &$-$69.75 & 0.23 \\
                &      &     & 0.46 &$-$69.43 & 0.62 \\
                &      &     & 1.19 &$-$45.90 & 0.30 \\
                &      &     & 2.27 &$-$45.30 & 0.62 \\
                &      &     & 3.46 &$-$44.50 & 0.31 \\
                &      &     & 2.70 &$-$44.45 & 0.57 \\
                &      &     & 5.53 &$-$43.98 & 0.18 \\
                &      &     &109.41&$-$43.60 & 0.28 \\
                &      &     &55.68 &$-$43.40 & 0.33 \\
                &      &     &105.34&$-$42.89 & 0.23 \\
                &      & RCP & 0.19 &$-$70.56 & 0.39 \\
                &      &     & 0.42 &$-$70.10 & 0.26 \\
                &      &     & 0.81 &$-$69.73 & 0.26 \\
                &      &     & 0.66 &$-$47.42 & 0.24 \\
                &      &     & 0.48 &$-$46.89 & 0.28 \\
                &      &     & 0.24 &$-$45.51 & 0.30 \\
                &      &     & 1.31 &$-$44.98 & 0.20 \\
                &      &     & 2.18 &$-$43.98 & 2.32 \\
                &      &     & 1.74 &$-$43.96 & 0.29 \\
                &      &     & 6.00 &$-$43.34 & 0.30 \\
                &      &     &39.90 &$-$42.83 & 0.29 \\
                &      &     &75.26 &$-$42.55 & 0.27 \\
                &      &     &11.56 &$-$42.03 & 0.38 \\
                & 6035 & LCP & 2.26 &$-$48.86 & 0.18 \\
                &      &     & 2.22 &$-$47.77 & 0.88 \\
                &      &     & 4.13 &$-$47.26 & 0.44 \\
                &      &     &34.20 &$-$45.62 & 0.26 \\
                &      &     &17.17 &$-$45.23 & 0.37 \\ 
                &      &     &30.25 &$-$44.35 & 0.58 \\ 
                &      &     &195.57&$-$43.37 & 0.36 \\ 
                &      &     &342.11&$-$42.99 & 0.21 \\ 
                &      &     & 54.75&$-$42.81 & 0.42 \\
                &      & RCP &13.60 &$-$46.97 & 0.31 \\
                &      &     &36.70 &$-$45.04 & 0.34 \\
                &      &     &16.07 &$-$44.46 & 0.37 \\
                &      &     &39.58 &$-$44.06 & 0.26 \\
                &      &     &52.55 &$-$43.52 & 0.25 \\
                &      &     &156.35&$-$43.06 & 0.17 \\
                &      &     &141.39&$-$42.75 & 0.91 \\
Sgr B2M         & 6030 & LCP & 0.05 &   53.63 & 8.87 \\
                &      &     & 0.08 &   65.06 & 8.71 \\
                &      &     & 0.05 &   65.73 & 0.55 \\
                &      &     & 0.21 &   70.64 & 0.24 \\
                &      &     & 0.69 &   72.21 & 0.22 \\
                &      & RCP & 0.05 &   52.98 & 7.70 \\
                &      &     & 0.08 &   64.93 & 9.94 \\
                &      &     & 0.05 &   65.35 & 0.54 \\
                &      &     & 0.49 &   70.21 & 0.23 \\
                &      &     & 0.16 &   71.58 & 0.12$^\mathrm{a}$ \\
                &      &     & 0.11 &   71.99 & 0.57 \\
                &      &     & 0.12 &   72.43 & 0.18 \\
                & 6035 & LCP & 0.02 &   50.59 & 3.46 \\
                &      &     & 0.30 &   62.05 & 0.19 \\
                &      &     & 0.08 &   62.58 & 1.08 \\
                &      &     & 0.13 &   63.58 & 0.56 \\
                &      &     & 0.04 &   64.00 & 5.30 \\
                &      &     & 0.46 &   65.02 & 0.31 \\
                &      &     & 0.08 &   65.89 & 0.93 \\
                &      &     & 0.07 &   66.48 & 0.25 \\
                &      &     & 0.54 &   67.38 & 0.51 \\
                &      &     & 0.12 &   68.90 & 0.97 \\
                &      &     & 1.09 &   70.65 & 0.40 \\
                &      &     & 1.18 &   71.04 & 0.38 \\
                &      &     & 4.33 &   72.23 & 0.32 \\
                &      & RCP & 0.05 &   48.83 & 0.39 \\
                &      &     & 0.03 &   50.17 & 3.75 \\
                &      &     & 0.04 &   60.90 & 1.22 \\
                &      &     & 0.27 &   62.38 & 0.61 \\
                &      &     & 0.22 &   63.26 & 0.74 \\
                &      &     & 0.37 &   64.62 & 0.38 \\
                &      &     & 0.16 &   65.88 & 1.79 \\
                &      &     & 0.63 &   66.99 & 0.51 \\
                &      &     & 0.18 &   68.92 & 1.40 \\
                &      &     & 1.30 &   70.26 & 0.27 \\
                &      &     & 1.78 &   70.63 & 0.44 \\
                &      &     & 2.18 &   71.62 & 0.32 \\
                &      &     & 1.22 &   72.25 & 0.23 \\
                &      &     & 1.85 &   72.66 & 0.35 \\
G5.886$-$0.393  & 6030 & LCP & 0.13 &    9.07 & 3.09 \\
                &      & RCP & 0.13 &    8.91 & 6.47 \\
                & 6035 & LCP & 0.13 &    8.53 & 6.79 \\
                &      &     & 0.14 &    9.73 & 0.80 \\
                &      & RCP & 0.22 &    8.46 & 7.56 \\
                &      &     & 0.14 &    9.76 & 0.90 \\
G10.624$-$0.385 & 6035 & \multicolumn{4}{c}{See \S \ref{sourcenotes}} \\
M17             & 6030 & LCP & 2.85$^\mathrm{b}$& 21.49 & 0.34 \\
                &      & RCP & 2.65$^\mathrm{b}$& 21.55 & 0.44 \\
                & 6035 & LCP &66.01$^\mathrm{b}$& 21.39 & 0.31 \\
                &      &     &61.21$^\mathrm{b}$& 22.58 & 0.23 \\
                &      &     & 5.43$^\mathrm{b}$& 23.54 & 0.34 \\
                &      & RCP&\nodata$^\mathrm{b}$&21.43 & 0.33 \\
                &      &    &\nodata$^\mathrm{b}$&22.58 & 0.32 \\
G28.199$-$0.048 & 6030 & LCP & 0.17 &   94.42 & 0.51 \\
                &      &     & 0.18 &   96.17 & 0.67 \\
                &      &     & 0.12 &   96.74 & 4.38 \\
                &      & RCP & 0.38 &   94.97 & 0.37 \\
                &      &     & 0.18 &   96.06 & 0.68 \\
                &      &     & 0.18 &   96.24 & 3.56 \\
                & 6035 & LCP & 1.05 &   94.51 & 0.59 \\
                &      &     & 3.56 &   96.07 & 0.30 \\
                &      &     & 1.05 &   96.32 & 1.00 \\
                &      &     & 0.21 &   96.61 & 3.92 \\
                &      &     & 0.13 &   97.79 & 0.25 \\
                &      & RCP & 1.59 &   94.90 & 0.63 \\
                &      &     & 3.13 &   96.12 & 0.66 \\
                &      &     & 0.18 &   96.86 & 4.40 \\
                &      &     & 0.38 &   97.02 & 0.62 \\
                &      &     & 0.14 &   98.29 & 0.43 \\
W48             & 6035 & LCP & 0.39 &   40.23 & 0.14 \\
                &      &     & 2.59 &   42.97 & 0.25 \\
                &      &     & 2.31 &   43.33 & 0.47 \\
                &      &     & 0.73 &   44.18 & 0.46 \\
                &      & RCP & 0.29 &   40.31 & 0.31 \\
                &      &     & 2.80 &   42.93 & 0.22 \\
                &      &     & 3.25 &   43.28 & 0.62 \\
                &      &     & 1.28 &   44.18 & 0.35 \\
G45.122$+$0.133 & 6030 & LCP & 0.18 &   53.89 & 0.65 \\
                &      &     & 0.08 &   54.51 & 0.64 \\
                &      & RCP & 0.14 &   53.51 & 0.50 \\
                &      &     & 0.11 &   54.15 & 0.85 \\
                & 6035 & LCP & 1.90 &   53.82 & 0.37 \\
                &      &     & 1.19 &   54.38 & 0.77 \\
                &      &     & 0.86 &   55.49 & 0.40 \\
                &      & RCP & 2.02 &   53.58 & 0.34 \\
                &      &     & 2.72 &   54.16 & 0.61 \\
                &      &     & 1.14 &   55.42 & 0.33 \\
W51 e/d         & 6030 & LCP & 0.30 &   52.33 & 0.33 \\
                &      &     & 0.89 &   52.95 & 0.28 \\
                &      & RCP & 0.25 &   52.73 & 0.43 \\
                &      &     & 1.00 &   53.35 & 0.30 \\
                & 6035 & LCP & 0.37 &   52.63 & 1.03 \\
                &      &     & 1.71 &   55.20 & 1.04 \\
                &      &     & 0.26 &   56.58 & 1.19 \\
                &      &     & 4.02 &   57.43 & 0.31 \\
                &      &     & 0.47 &   63.05 & 0.39 \\
                &      &     & 0.95 &   63.92 & 0.20 \\
                &      & RCP & 0.64 &   52.95 & 1.01 \\
                &      &     & 1.39 &   55.46 & 1.24 \\
                &      &     & 1.16 &   55.67 & 0.70 \\
                &      &     & 0.45 &   56.84 & 0.51 \\
                &      &     & 5.58 &   57.74 & 0.36 \\
                &      &     & 0.50 &   63.17 & 0.33 \\
                &      &     & 0.55 &   64.15 & 0.31 \\
K3-50           & 6030 & LCP & 0.29 &$-$19.10 & 0.33 \\
                &      & RCP & 0.25 &$-$19.80 & 0.30 \\
                & 6035 & LCP & 1.05 &$-$19.29 & 0.39 \\
                &      &     & 0.46 &$-$18.60 & 0.60 \\
                &      & RCP & 1.04 &$-$19.78 & 0.44 \\
                &      &     & 1.59 &$-$18.99 & 0.34 \\
ON 1            & 6030 & LCP & 0.38 &    0.69 & 0.27 \\
                &      &     & 0.10 &    1.30 & 0.36 \\
                &      &     & 2.57 &   14.17 & 0.29 \\
                &      & RCP & 0.18 & $-$0.37 & 0.45 \\
                &      &     & 3.64 &   13.80 & 0.26 \\
                & 6035 & LCP & 0.12 & $-$0.76 & 0.19 \\
                &      &     & 0.42 & $-$0.13 & 0.46 \\
                &      &     & 1.41 &    0.49 & 0.26 \\
                &      &     & 0.26 &    1.10 & 0.27 \\
                &      &     & 0.15 &    1.68 & 0.32 \\
                &      &     & 0.25 &    2.11 & 0.39 \\
                &      &     & 0.14 &    5.47 & 0.39 \\
                &      &     & 0.09 &    7.74 & 0.21 \\
                &      &     & 0.26 &   12.59 & 0.27 \\
                &      &     & 2.06 &   14.02 & 0.65 \\
                &      &     & 3.09 &   14.52 & 0.21 \\
                &      &     & 0.68 &   14.88 & 0.37 \\
                &      &     & 1.96 &   15.44 & 0.26 \\
                &      & RCP & 1.21 & $-$0.71 & 0.26 \\
                &      &     & 1.69 & $-$0.24 & 0.32 \\
                &      &     & 0.53 &    0.47 & 0.12$^\mathrm{a}$ \\
                &      &     & 0.29 &    1.38 & 0.26 \\
                &      &     & 0.34 &    1.82 & 0.32 \\
                &      &     & 0.47 &    5.53 & 0.39 \\
                &      &     & 5.56 &   13.81 & 0.38 \\
                &      &     &15.91 &   14.47 & 0.30 \\
                &      &     &$\approx 4^\mathrm{c}$&15.16&0.13\\
DR 20           & 6030 & LCP & 1.43 &$-$10.92 & 0.23 \\
                &      & RCP & 1.18 &$-$11.19 & 0.22 \\
                & 6035 & LCP & 1.96 &$-$11.02 & 0.30 \\
                &      &     & 6.26 & $-$9.64 & 0.16 \\
                &      & RCP & 1.85 &$-$11.18 & 0.37 \\
                &      &     & 1.92 & $-$9.87 & 0.33 \\
W75 N           & 6035 & LCP & 0.44 &    7.10 & 0.34 \\
                &      &     & 4.59 &    7.65 & 0.27 \\
                &      &     & 1.78 &    9.41 & 0.26 \\
                &      & RCP & 0.58 &    6.88 & 0.38 \\
                &      &     & 1.54 &    7.52 & 0.21 \\
                &      &     & 4.06 &    8.05 & 0.35 \\
                &      &     & 2.01 &    9.67 & 0.32 \\
W75 S           & 6035 & LCP & 0.97 & $-$8.79 & 0.12$^\mathrm{a}$\\
                &      &     & 0.20 & $-$2.24 & 0.21 \\
                &      &     & 1.20 &    3.46 & 0.30 \\
                &      & RCP & 0.12 & $-$8.79 & 0.40 \\
                &      &     & 0.20 & $-$2.37 & 0.30 \\
                &      &     & 0.55 &    3.20 & 0.31 \\
                &      &     & 0.21 &    3.76 & 0.52 \\
LDN 1084        & 6035 & LCP & 0.19 &$-$62.84 & 0.18 \\
                &      &     & 1.24 &$-$61.51 & 0.23 \\
                &      & RCP & 0.20 &$-$62.60 & 0.29 \\
                &      &     & 1.04 &$-$61.46 & 0.29 \\
NGC 7538        & 6030 & LCP & 0.10 &$-$59.12 & 1.18 \\
                &      & RCP & 0.09 &$-$59.14 & 1.06 \\
                & 6035 & LCP & 1.64 &$-$59.34 & 0.40 \\
                &      &     & 0.20 &$-$58.89 & 0.32 \\
                &      &     & 0.06 &$-$58.74 & 3.16 \\
                &      &     & 0.09 &$-$56.91 & 0.29 \\
                &      & RCP & 1.71 &$-$59.34 & 0.48 \\
                &      &     & 0.11 &$-$58.83 & 0.38 \\
                &      &     & 0.08 &$-$58.70 & 2.95 \\
                &      &     & 0.16 &$-$56.84 & 0.29 \\
\hline
\end{longtable}
\begin{list}{}{}
\item[$^{\mathrm{a}}$]FWHM is comparable to channel width.

\item[$^{\mathrm{b}}$]Saturation of the autocorrelator prevented
                accurate determination of flux densities.  See Sect.\
                \ref{sourcenotes} for further details.
\item[$^{\mathrm{c}}$]Untapered spectrum is contaminated by ringing,
                but there is insufficient spectral resolution to
                achieve a good fit of the maser parameters after
                Hanning weighting.
\end{list}

\end{document}